\documentclass[prb,showpacs,preprintnumbers,twocolumn]{revtex4}
\usepackage[dvips]{graphicx}
\usepackage{amsmath,amsfonts,amssymb,txfonts}
\usepackage{epsfig}
\begin{document}

\title{Free surface instability in a confined suspension jet}
\author{Alejandra Alvarez$^{1,2}$, Eric Clement$^3$ and Rodrigo
Soto$^{1,4}$}
\email{rsoto@dfi.uchile.cl}
\affiliation{$^1$ Departamento de F\'{\i}sica, FCFM, Universidad de Chile,
  Casilla 487-3, Santiago, Chile.\\
$^2$ Instituto de Innovaci\'on en Mineria y Metalurgia, IM2, avda. del
Valle 738, Huechuraba, Santiago, Chile.\\
$^3$ L.M.D.H. - PMMH ESPCI. 10, rue Vauquelin. 75231 Paris Cedex 05,
France\\
$^4$ Departamento de F\'{\i}sica Aplicada I (Termolog\'\i a),
Facultad de Ciencias F\'{\i}sicas, Universidad Complutense,
28040 Madrid, Spain}

\date{\today}

\begin{abstract}

A jet of non-Brownian particles confined in a thin cell and driven by
gravitational force is studied both numerically and theoretically. We
present a theoretical scheme aimed to describe such a system in the Stokes
regime. We focus on the dynamics of the interface between the suspension and
the pure fluid. Numerical simulations solving Newton's equations for all
particles show that the jet free surface becomes unstable: the fastest
growing modes at small sizes coarsen up to the largest structures reaching
the jet lateral scale. In the bulk, structural waves develop and travel at
slightly slower speed than the jet average fall. An analytical model, based on
hydrodynamic-like equations for the suspension, is derived and predicts the
development of the interfacial instability. It captures in essence, the
collective effects driving the interface destabilization i.e. the long
range hydrodynamic interactions coupled with the abrupt interface, and no
analogous to a surface tension is found.

\end{abstract}


\maketitle

\section{Introduction}

Understanding the dynamics of 3D non-Brownian suspension in the low
Reynolds number regime is a long lasting and difficult issue. Long-range
hydrodynamics forces create a complex collective particle dynamics
\cite{Happel65, refe_SD} and up to now, no rigorous closed-form formulation
of the problem exists at moderate densities. For example, difficulties
remains such as to to explain particle sedimentation, dispersion and mixing
in a finite size container (see \cite{Ladd02} and references inside).
Recently,
the miniaturization of hydrodynamics devices \cite{lubrica, Stone04},
necessary to develop microfluidic apparatus performing at low Reynolds
number, mixing or separation, has brought to the front the importance of
boundary confinement of suspensions. In fact, numerous devices bear the
form of narrow channels or Hele-Shaw cells with one direction reaching only
few particle sizes \cite{Weigl99}. Note that experimentally quasi-2D
non-Brownian suspensions have also been studied in the case where the cell
gap was almost the particle size \cite{Rouyer1999, Santana01,
cui2004}. In this case, the existence of anti-drag correlations due to
fluid recirculation around each grains were identified \cite{cui2004}. 

To get a better description of mixing, an important basic issue is to
capture the evolution of an interface between a suspension and a
particle-free fluid (the pure fluid). This problem is also related to many
studies done to understand miscible interfaces dynamics, either from two
different fluids \cite{petitjean99} or a fluid in contact with a suspension
falling under gravity \cite{refe8, refe11, refe9}.  

Here, we describe the evolution of a jet of suspended particles in a thin
cell driven by gravitational forces. The suspension is non Brownian and the
hydrodynamic forces between particles are obtained in the Stokes regime.
Numerical simulations that solve the Newton equations for all particles
follow the evolution of the jet free surfaces. We present an analytical
model, based on hydrodynamic-like equations for the suspension which is
able to predict the development of the instability. We present and extend a
scheme that was developed recently, in the context of a 2D suspension
cluster falling down \cite{hele_suspe} and also presented briefly in
ref.\cite{hele_suspe_jet} in the context of a jet. The relative simplicity
of the model, focusing on the actual effect of long range hydrodynamic
forces, allows to discuss the dominant physical features of the interface
dynamics.

\section{Model} 

We consider a system of $N$ solid particles that move through an
incompressible Newtonian fluid of viscosity $\eta$. The fluid is confined
between two parallel plates separated at a distance $2d$ in the $z$
direction, being infinite in the other two directions. To simplify the
computation of the hydrodynamic interaction forces, and to focus on the
effect of the long-range nature of them, we consider cylindrical
particles of height $L$ (slightly smaller than $2d$) and radius $\sigma$
that have planar motion only. Particles are thin, i.e. $2d\ll\sigma$ and
their mass is $m$.

The hydrodynamic forces between the cylinders in the confined geometry
have been computed in Ref. \cite{hele_suspe} in the Stokes regime 
for a dilute suspension.  First,
the force over the $i$-th particle has a drag component given by
$-\frac{m}{\tau_1}\vec u_{i}$, where $\vec u_{i}$ is the in-plane
velocity of particle $i$ and $\tau_1=md/\pi\sigma^2\eta$ is the relaxation
time of a single particle. Also, the presence of the other particles
induce hydrodynamic forces over each other. When particles are far
apart, the force on $i$ due to the presence of $k$ is
\begin{eqnarray}  
\vec F_{ik}^{F}&=& -\frac{m}{8\tau_1} \varmathbb{K}
(\vec
R_{ki})\vec u_{k} \label{farforce}
\end{eqnarray}
where $\vec R_{ki}=\vec R_{i}-\vec R_{k}$ is the relative distance between
particles, being $\vec R_{i}$ the position of the center of mass of the $i$-th
particle and the tensor $\varmathbb{K}$ is given by
\begin{eqnarray}
\varmathbb{K}(\vec{R})=
(\sigma/R)^{2}(\varmathbb{I}-2\vec{R}\vec{R}/R^2)\label{kernel} 
\end{eqnarray}
This force depends both on the
direction of the velocity $\vec{u}_k$ and on the relative distance
$\vec{R}_{ik}$. When $\vec{u}_k$ is parallel to $\vec{R}_{ik}$, the
interaction force on $i$ is parallel to $\vec{u}_k$ (drag), and if
$\vec{u}_k$ is
perpendicular to $\vec{R}_{ik}$, the force turns out to be in an opposite
direction to $\vec{u}_k$ (antidrag).

On the other hand, when particles are close to each other, lubrication
forces appear producing the net force on particle $i$ due to
the presence of $k$ ~\cite{hele_suspe}
\begin{eqnarray} 
\vec F_{ik}^N&=&2\pi d\eta
\frac{1}{\sqrt{\epsilon}}\frac{1}{R_{ik}^2}
\left(R_{ik}^2 \varmathbb{I} -\vec{R}_{ik}\vec{R}_{ik} \right)
\vec{u}_{ik} \nonumber\\
&&+
\frac{3}{2} 2\pi d \eta\frac{d^2}{\sigma^2}  \frac{1}{\epsilon} \frac{\vec
u_{ik}\cdot\vec R_{ik}}{R_{ik}^2} \vec R_{ik} \label{lubrication}
\end{eqnarray}
where $\epsilon=R_{ik}/\sigma-1$ is the gap between the particles. The force
depends on the relative velocities between the pair and it
respects the action-reaction principle. Therefore it does not change the
total momentum of the pair, but reduces the relative velocity.

The cutoff for using either expression (\ref{farforce}) or
(\ref{lubrication})
is rather arbitrary. We adopt the convention that when the pair is closer
than $R_{\rm lubr}$, the lubrication force~(\ref{lubrication}) is used, when
the distance is larger than $R_{\rm far}$, the far force~(\ref{farforce}) is
used, and in between a linear interpolation between the two is computed. The
result is the interaction force
$\vec{F}^I_{ik}$.

There is also an extra drag force produced by the flow between the particles and
the plates \cite{hele_suspe}. This force can be added to the other drag
force, simply modifying the prefactor by a value given by the
particular experimental setup. In order to simplify the analysis, in what
follows we will disregard the presence of this term but in case of being
relevant to some experimental configuration, it can be trivially
included.

In summary, the dynamical equations for the suspended particles are
\begin{eqnarray} 
m\frac{d\vec u_i}{d t}&=&-\frac{m}{\tau_1} \vec u_{i}+
\sum_k
\vec F_{ik}^{I} +m\vec g  \label{ecsmov}
\end{eqnarray}
that constitute a complex system of equations that will be solved
numerically. 

Although the tensor (\ref{kernel}) diverges at close distances, it has the
remarkable property that, its integral over the direction of $\vec{R}$
vanishes. This implies that if a particle is surrounded by an homogeneous
distribution of suspended particles, all moving with equal velocities, the
far-field contribution to the hydrodynamic force vanishes. 

The long range nature of the forces forces between particles makes it
computationally expensive to perform direct numerical simulations where all
pairs of forces are computed.  However, the expressions (\ref{farforce}) and
(\ref{kernel}) for the far-field contribution to the hydrodynamic forces,
suggests that a mean field approximation can be done. In fact, if particles
in a region move with similar velocities they exert similar forces to a
target particle; it is natural then to group them altogether and compute the
total force made by the group on any target particle. We define the coarse
grained current $\vec{J}$ on a cell $c$ as

\begin{equation}
\vec{J}_c = \frac{1}{\Delta S}\sum_{i\in c} \vec{u}_i
\end{equation}
where $\Delta S$ is the area of each cell.

Then, if the system is nearly homogeneous in each cell, the far field
contribution to the force (\ref{farforce}) can be approximated by
\begin{equation}
\vec F_{i } = 
-\frac{m}{8\tau_1} \sum_{c}\Delta S \varmathbb{K}(\vec{R}_i-\vec{R}_c)
\vec{J}_c  \label{force.meanfield}   
\end{equation} 
where the sum runs on all
cells, $\vec{R}_c$ is the position of the center of each cell. This
expression allows to study systems of many particles at reasonable cost. 
Note that the vanishing integral of $\varmathbb{K}(\vec{R})$ over the
angles implies that~(\ref{force.meanfield}) can be applied down to
neighboring cells without producing divergences, but 
the mean field approximation can be inaccurate for near cells. Local
inhomogeneities are not completely taken into account by
the mean field and unrealistic vanishing forces can be obtained. To
solve this problem, the mean field
approximation of the far field force is applied only to cells that are
separated by a distance larger than $R_{\rm mf}$, otherwise the direct
summation on the pair of particles is performed.
Finally, the values chosen for performing the simulations are $R_{\rm
lubr}=1.3\sigma$, $R_{\rm far}=2.0\sigma$, $R_{\rm mf}=5.66 \sigma$.
This numerical method was shown to give accurate results in the study of
the spreading of a falling cluster \cite{hele_suspe}.

\section{Simulations of a falling jet}

The long range forces (\ref{farforce}) vanish when a particle is surrounded
by and homogeneous medium, but in presence of inhomogeneities the force is
finite. The effect of inhomogeneities is clearly seen when there a
separation line between a region with suspended particles and a region
without them; it was shown in \cite{hele_suspe}, that the force is enhanced
if the separation line is curved,  being possible to produce
instabilities. As our description is for the
dynamics of the suspended particles, and not for the surrounding fluid, it
is sensible to call {\em free surface} the separation line between a region
with suspended particles and the pure fluid.

In order to study the effect of the long range forces on the free surfaces
and, in particular, if they can lead to unstable motion, we consider a
falling jet of particles immersed in a fluid.

A jet of falling particles is studied numerically. Initially, a
jet of $N=24000$ particles is placed randomly at rest in a
rectangle of width $L_x=90\sigma$ and height $L_y=600\sigma$. They 
fall down due to the action of a gravitational field pointing in the
positive $y$ direction. To mimic an infinity jet
the vertical direction is periodic and the force computation uses the
minimum image convention \cite{Allen}. 
Units are chosen such that the particle diameter $\sigma$, particle mass
$m$, and the relaxation time $\tau_1$ are set
to one. The gravitational force is $mg=1.0$ and therefore the 
limiting velocity for a single particle $V^{\infty}_1=g\tau_1=1.0$ (see
Eq. (\ref{ecsmov})).

\begin{figure}[h]
\begin{center}
\includegraphics[angle=0,width=0.85\columnwidth]{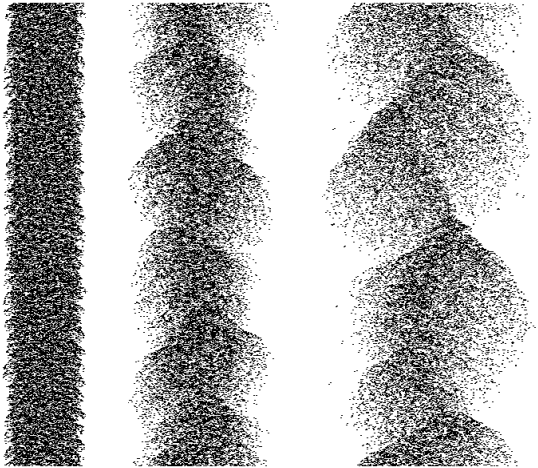}
\caption{
Numerical simulation of a suspension of $N=24000$ particles falling by
the action of gravitational field. From the left to the right:
$t=200$, $t=2000$, and $t=4000$. Units are described in the text.}
\label{Fig.interface1}
\end{center}
\end{figure}

Fig.~\ref{Fig.interface1} shows three successive snapshots of the
jet. They show that the free surfaces become unstable,
showing the appearance of waves. At the beginning the
surface waves are characterized by short wavelengths but later as the
amplitude of the perturbations growth, the characteristic wavelengths grow
also. A coarsening process is developed leading to larger structures. Once
the size of this structures is comparable to the jet width, interactions
between the two surfaces are observed and in-phase surface oscillations are
obtained. The surface waves are also accompanied by modulations
of the particle density.

\begin{figure}[h]
\begin{center}
\includegraphics[angle=0,width=0.95\columnwidth]{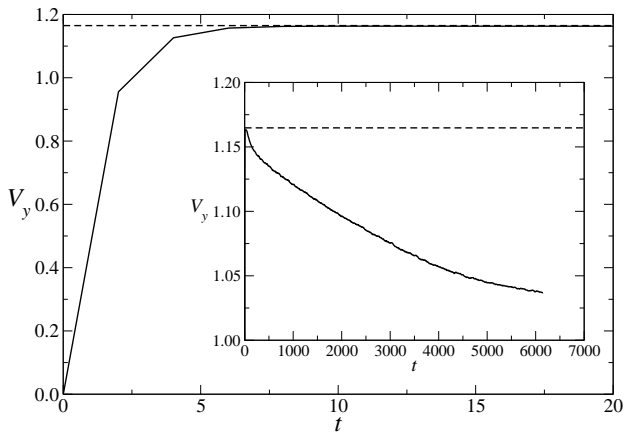}
\caption{
Instantaneous average of the vertical velocity as a
function of time (continuous line). The dashed line is the asymptotic
theoretical value of the jet if it does not deform, $V_{\rm
jet}^{\infty}=1.165$. Inset: evolution for longer times.}
\label{Fig.velCM}
\end{center}
\end{figure}

Fig. \ref{Fig.velCM} shows the average vertical velocity of the falling
jet as a function of time. It is seen that the jet rapidly gets an
asymptotic velocity that is larger than the one for a single isolated
particle. At much larger times, as the
jet develops large structures and the particles separate, the falling
velocity decreases and approaches $V_1^{\infty}=1.0$.

\begin{figure}[h]
\begin{center}
\includegraphics[width=0.6\columnwidth]{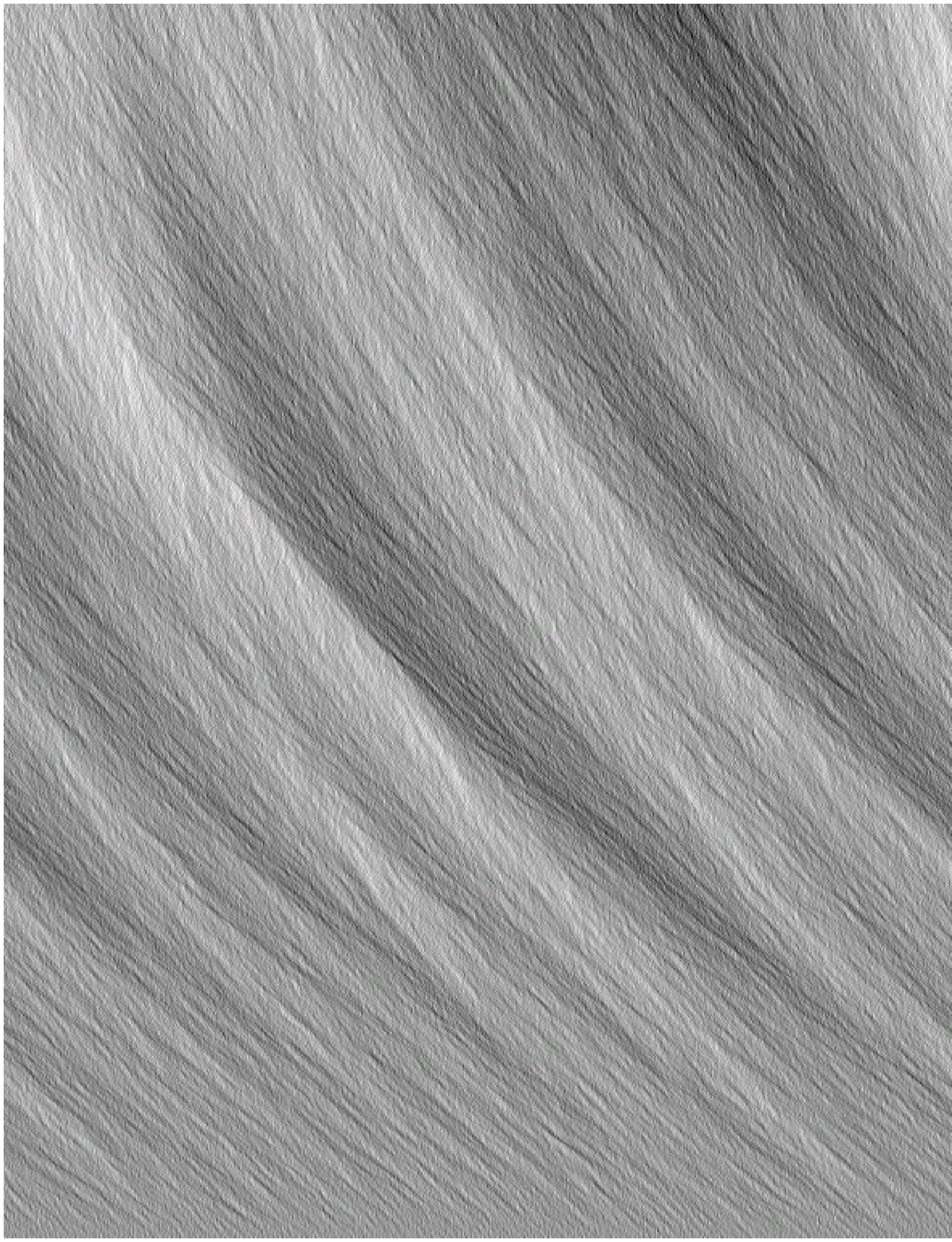}
\caption{Spatiotemporal diagram of the evolution of the 
$x$-integrated density $N_x(y,t)$ in the reference frame of the
falling jet. Time is on the vertical axis
and increasing upward. The horizontal axis is the $y$ coordinate, with
periodic boundary conditions, and the
gravitational acceleration points to the right. The gray scale is
proportional to density, with lighter regions representing denser regions
of the system, the minimum and maximum represented densities are
$N_x^{\rm min}=11$ and $N_x^{\rm max}=82$,
respectively, and the average is $N_x^{\rm avg}=N/L_y=40$.
The simulation parameters are the same as in Fig. \ref{Fig.interface1} and
the total simulation time is $t=6152$.}
\label{Fig.xt}
\end{center}	
\end{figure}

To describe in more detail the coarsening process, the $x$-integrated
density $N_x(y,t)$ is studied, computed as the number of particles in a
horizontal bin of height $\sigma$ centered at $y$. In Fig.
\ref{Fig.xt} a spatio-temporal plot $N_x(y,t)$ is presented in
the reference frame falling with the same instantaneous velocity of the
jet. It is seen that structures grow
in time showing the coarsening process. Also, the structures propagate in
the negative $y$ direction; as the plot is presented in the reference frame
of the falling jet it indicates that the perturbations fall at a slightly
smaller velocity than the jet.

\section{Global model}
The behavior observed in the numerical simulations suggest an
instability like the observed when two immiscible fluids in contact move
with a relative velocity (Kelvin-Helmholtz instability) \cite{chandra}. 

To describe the appearance of the surface instability, we build a global
model,
similar to hydrodynamic equations. We consider the particle number
density $n(\vec{R})$ and the particle mean velocity $\vec{V}(\vec
R)$. The average force density over the suspension, produced by the drag
force, the far force contribution (\ref{farforce}), and the gravity
acceleration is 
\begin{eqnarray}
\vec F(\vec R) &=&   -\frac{1}{\tau_1} \vec J(\vec R)
-\frac{1}{ 8\tau_1}  n(\vec R)  \int d\vec R'  \varmathbb{K} (\vec
R-\vec R') \vec J(\vec R') +n(\vec R)m\vec g \nonumber
\label{forcedensity}\\
\end{eqnarray}
with $\vec{J}=m n \vec{V}$ is the mass current density.

In a Euler-like global model as the one proposed, the lubrication force
contribution can be neglected because its effect is to reduce velocity
fluctuations, but it does not modify the mean velocity as it only affects
the relative velocity. Eventually, its effect would be to produce an
additional effective viscosity as obtained in kinetic theory
\cite{KT}.

Therefore a global equation of motion for the suspension can be written as
\begin{eqnarray}
&&\frac{\partial n}{\partial t}+\nabla\cdot
\left(n \vec V\right)=0\label{mass.cons}\\
&&n m\left(\frac{\partial \vec V}{\partial t}+\vec
V\cdot\nabla \vec V\right)=\vec F\label{momentum.cons}
\end{eqnarray}

Before analyzing the development of the instabilities, let's
consider
the evolution of a unperturbed homogeneous jet. For a thin jet of width
$L_x$ and height $L_y$, with periodic boundary
conditions in $y$, if the particle current is assumed to be homogeneous
$\vec{J}(\vec R,t) = J_{\rm jet}(t)\hat{y}$, the integral term in
(\ref{forcedensity}) simplifies (see Eq. (\ref{IntKFinito})) and we have
that 
\begin{eqnarray}
&&\int d\vec R'  \varmathbb{K} (\vec
R-\vec R') \vec J_{\rm jet} \nonumber \\
&&\quad = 2\sigma^2\left[\arctan(L_y/L_x)-\arctan(L_x/L_y)\right] J_{\rm
jet}\hat{y}
\end{eqnarray}
independent of $\vec R$.
Therefore Eqs. (\ref{mass.cons}) and (\ref{momentum.cons}) admit a
solution of an homogeneous falling jet with velocity
\begin{equation}
V_{\rm jet}(t) = g \tau_{\rm
jet}\left(1-e^{-t/\tau_{\rm jet}}\right) \label{velasympt}
\end{equation}
where the jet relaxation time is
\begin{equation}
\tau_{\rm jet} = \frac{\tau_1}
{1-n_0\sigma^2\left[\arctan(L_y/L_x)-\arctan(L_x/L_y)\right]/4 }
\label{relaxtime}
\end{equation}
and $n_0$ is the jet number density.

Note that the asymptotic velocity as the relaxation time is different
from the one of a single falling particle $V^{\infty}_1=g\tau_1$. This
effect is due to the long
range hydrodynamic interactions. A peculiar feature is that the
modification of the falling velocity depends only on the system shape,
but not on its size. A similar phenomena was observed in the case of a
falling circular cluster \cite{hele_suspe}.
In fact, for a thin jet ($L_x<L_y$) most of the relative
distances between particles are parallel to the falling velocity, giving
rise to a positive hydrodynamic forces between particles. That is
particles exerts a positive drag, in the direction of the motion, 
increasing the falling velocity. On the other hand, if the jet were wide
($L_x>L_y$) most of the relative
distances between particles would be perpendicular to the falling
velocity and the dominant contribution of the hydrodynamic forces would be
the antidrag, reducing the jet falling velocity.
In the case of the jet we are considering the theoretical values for the
falling jet velocity and relaxation time are $V_{\rm
jet}^{\infty}=1.165$ and $\tau_{\rm jet}=1.165$, that are larger than
the values for a single particle. In Fig.
\ref{Fig.velCM} it is seen that the jet rapidly gets this velocity and
afterward, when it starts deforming, the velocity slowly decreases
approaching $V_1^{\infty}=1.0$. This long time value is consistent 
with Eqs. (\ref{velasympt}) and (\ref{relaxtime}), when the jet
density decreases. An exponential fit to the early time evolution of the
velocity of the jet gives the same relaxation time as predicted.

\begin{figure}[h]
\begin{center}
\epsfclipon\includegraphics[angle=0,width=0.6\columnwidth]{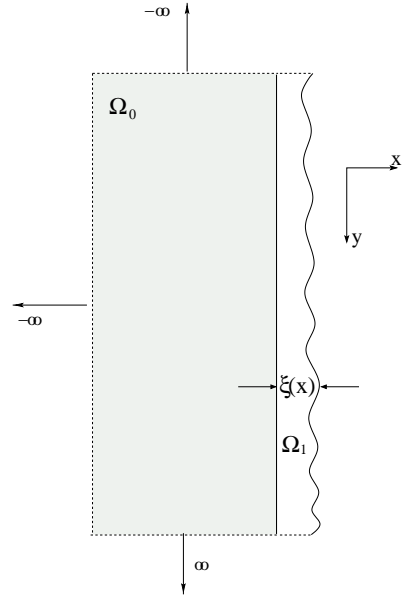}

\caption{Left: Sketch of the semi - infinity suspension. $\Omega_o$ is the
unperturbed domain, $\Omega_1$ is the perturbed domain, $\xi$ is the
perturbation in the $x$ direction.}

\label{Fig.sketch}
\end{center}
\end{figure}

In presence of a free surface, an additional equation must be added to
describe the evolution of the free surface position $\xi(y,t)$ (see Fig.
\ref{Fig.sketch}). The mean suspension velocity at the surface must be equal
to the surface velocity \cite{chandra}. 
\begin{eqnarray}
\frac{\partial \xi}{\partial t} +V_y\Big
|_{x=\xi} \frac{\partial
\xi}{\partial y}&=& V_x\Big|_{x=\xi} \label{eq.xi}
\end{eqnarray}

\section{Linear Stability Analysis}

The stability of the free surface of the jet is analyzed using the global Eqs.
(\ref{mass.cons}), (\ref{momentum.cons}), and (\ref{eq.xi}). To
simplify the problem and assuming that at
the beginning the two free surfaces do not interact strongly, we will
consider the
simpler case of a single free surface, limiting a semi-infinite homogeneous
suspension in the $x<0$ region, of density $n_0$. In this geometry the
jet falling velocity is
$\vec{V}_0=g\tau_1\hat{y}/(1+\beta_0/4)$, where $\beta_0=\pi
\sigma^2 n_0/4$ is the area fraction of the suspension (see Eq.
(\ref{IntKSemiInf})). Note that in this geometry $V_0<V_1^\infty$. 

We consider small perturbations of the free surface. Introducing a formal
parameter $\epsilon\ll 1$, the fields are written as $\vec V(\vec
R)=\vec V_0+\epsilon \vec V_1(\vec R)$, $n=n_0+\epsilon n_1(\vec
R)$, and $\xi=\epsilon \xi_1(y)$. The domain $\Omega$ over which the
integration
in (\ref{forcedensity}) must be done can be split into the initial
domain
$\Omega_0:x\leq0$, plus a perturbed domain $\Omega_1: 0<x\leq \xi(y,t)$ (see
Fig.~\ref{Fig.sketch}). Keeping terms up to linear order in $\epsilon$, the
equations read
\begin{eqnarray}
&&\frac{\partial n_1}{\partial t}+n_0\nabla\cdot \vec V_1=0\\
&&\frac{\partial \vec V_1}{\partial t} +\vec V_0\cdot
\nabla \vec V_1= 
-\frac{1}{\tau_1}\vec V_1(\vec R)   
- \frac{1}{2\tau_1}\frac{n_1}{4}\varmathbb{K}(\hat y) \vec
J_0\nonumber\\
&&\qquad - \frac{1}{2\tau_1} n_0\int_{\Omega_0} d\vec R'  \varmathbb{K}
(\vec R-\vec R') \vec J_1(\vec R')\nonumber\\ 
&&\qquad -\frac{1}{2\tau_1} n_0\int_{-\infty}^{\infty}dy' \varmathbb{K}
(\vec R-y'\hat y) \vec J_0 \xi(y')\label{momentum.cons2} \\
&&\frac{\partial \xi}{\partial t} +V_{0y}\Big|_{x=0}
\frac{\partial
\xi}{\partial y}=V_{1x}\Big|_{x=0} 
\end{eqnarray}
The deformed surface has a long range effect of the fluid
motion, as reflected by the last term in Eq. \ref{momentum.cons2}.

This integro-differential system of equations does not have an evident
analytic solution. However, an estimation of the instability modes can be
obtained as follows. A Fourier expansion of the fields $A_1(\vec
R,t)=\hat{A}_1(\vec k,t)e^{i\vec k\cdot \vec R}$ is performed. As the
Fourier basis is not a solution of the system (particularly because of the
integral terms in (\ref{momentum.cons2})), we obtain an estimation
by evaluating the equations at $x=0$. Defining dimensionless variables
$\rho=\hat{n}_1/n_0$,
$\vec{U}=\widehat{\vec{V}}_1/(g\tau_1)$, 
$\varphi=\hat{\xi}/(g\tau_1^2)$,
$\vec{q}=(g\tau_1^2)\vec k$, and $s=t/\tau_1$, 
the equations read

\begin{eqnarray}
&&\frac{\partial \rho}{\partial s}=-i\vec q \cdot \vec
U -i\vec q\cdot \vec U_0 \rho\label{eq1adim} \\
&&\frac{\partial \vec U}{\partial s} = -\vec U - i(\vec q\cdot \vec U_0)
\vec U - \varmathbb{Q}_1 (\vec U + \rho \vec U_0) - \varmathbb{Q}_2 \vec
U_0 \varphi \label{eq2adim} \\ 
&&\frac{\partial \varphi}{\partial s} = U_x -i(\vec q \cdot \vec
U_0) \varphi \label{eq3adim} 
\end{eqnarray}
with the following defined tensors (see Eq. (\ref{FourierKy}))
\begin{eqnarray}
\varmathbb{Q}_1 &=& \frac{\beta_0}{2} \left\{
\left(\begin{array}{lr}1 & 0\\ 0 & -1\end{array}\right) 
+ \frac{|q_y|}{i q_x+ |q_y|}
\left(\begin{array}{rr}-1 & i\\ i & 1\end{array}\right)
\right\}\\
\varmathbb{Q}_2 &=& \frac{\beta_0}{2} |q_y|
\left(\begin{array}{rr}-1 & i\\ i & 1\end{array}\right)
\end{eqnarray}
We recall the the dimensionless jet falling velocity is $\vec{U}_0 =
\hat{y}/(1+\beta_0/4)$

The tensors  $\varmathbb{Q}_1$ and $\varmathbb{Q}_2$ capture the effect of
the long range forces produced by the perturbations, when they are
integrated in the semi-infinite volume. $\varmathbb{Q}_1$ describes the
interaction of the perturbations with the bulk and $\varmathbb{Q}_2$ the
interaction with the free surface. Remarkably, $\varmathbb{Q}_2$ is
proportional to $q$ and not to $q^2$, therefore the effect of a curved
surface cannot be described in terms of a surface tension. Note also that
$\varmathbb{Q}_1$ is independent of the magnitude of $\vec{q}$ and depends
only on its direction. This fact has an important consequence because in the
limit of perturbations of small wave vectors the effect of the hydrodynamic
interactions does not vanish. In fact in the limit $q\to 0$ the linear
system of equations (\ref{eq3adim}) has a Jordan-block structure and admits
a solution of the form (at first order in $\beta_0$)

\begin{eqnarray}
\rho &=& 2 A U_0^{-1} (-\cos\phi+i|\sin\phi|)\\
U_x  &=& A \beta_0 |\sin\phi|\\
U_y  &=& -A \beta_0 \cos\phi\\
\varphi&=& A \beta_0 |\sin\phi|\, t
\end{eqnarray}
where $A$ is an arbitrary coefficient given by the initial
condition and $\phi$ is the angle between the wave
vector $\vec q$ and the $\hat{x}$ direction. The solution shows a linear,
instead of exponential, increase in time of the surface oscillations.

For larger values of $q$ the system of equations looses its Jordan-block
structure and solutions in the form $\exp(\lambda s)$ are looked for. The
eigenvalues $\lambda$ for $q\ll 1$, $\beta_0\ll 1$, and
$0<\phi<\pi$ ($\sin\phi>0$) are
\begin{eqnarray}
\lambda_1&=& -i U_0 q (1+\beta_0/4)\sin\phi +\frac{\sqrt{3}}{4} \beta_0
U_0 q\sin\phi + O(q^2)\\
\lambda_2&=& -i U_0 q (1+\beta_0/4)\sin\phi -\frac{\sqrt{3}}{4} \beta_0
U_0 q\sin\phi +O(q^2)\\
\lambda_3&=& -1-\frac{\beta_0}{2}(\cos\phi+i\sin\phi) + O(q)\\
\lambda_4&=& -1+\frac{\beta_0}{2}(\cos\phi+i\sin\phi) + O(q)
\end{eqnarray}
and similar results for $\pi<\phi<2\pi$ ($\sin\phi<0$).

Two of the eigenvalues, $\lambda_3$ and $\lambda_4$ have negative real
parts
for small $q$ and therefore correspond to damped motion. However, the real
part of $\lambda_1$ ($\lambda_2$ for $\pi<\phi<2\pi$) is
positive. Therefore an instability is predicted. The
analysis shows that the system becomes unstable for any strength of the
gravitational force and, coming back to the original units, the instability
rate is directly proportional to $V_0 k$. In the limit $k\ll 1$ the real
parts of $\lambda_1$ and
$\lambda_2$ are proportional to the wavevector. It is reasonable to expect
that
a more detailed model, that includes terms proportional to gradients of
$\vec{V}$ or the viscous effect due to the lubrication forces, will
produce that for high enough wave vectors, the real part of the eigenvalues
become negative again. Hence, it is expected that the system is
unstable for a range of wave vectors going from zero to a finite value,
showing the coarsening process observed in the simulation
\cite{coarsening}.

The eigenvalues $\lambda_1$ and $\lambda_2$ have also an imaginary part.
The
dominant one with respect to $k$ and $\beta_0$ is $\lambda_{1I}=-iq
U_0(1+\beta_0/4)\sin\phi$. It corresponds to a wave in the form
$\exp(iq\sin\phi(y-U_0(1+\beta_0/4)t)+iq\cos\phi x)$, that propagates in the
$+y$ direction, with phase velocity
$U_0(1+\beta_0/4)=1$. That is, the structures propagate at the same velocity
than a single falling particle. This phase velocity is larger than the jet
falling velocity by a factor $O(\beta_0)$. As a consequence, the unstable
structures should move faster than the jet. In the simulations, we actually
find the opposite: the structures move slower than the jet. This change of
character can be due to the different geometries as they are considered,
as it also happens to the jet falling velocity that in one case is
faster than $V_1^\infty$, while in the other case was found slower. 

The linear stability analysis for a jet with two free surfaces as in
the simulations is much more involved. One extra equation must be
added and the long range interactions in this geometry give more complex
expressions, as can be seen in the case of the falling velocity.
Nevertheless, we expect that the analysis presented here,
showing that the surface is unconditionally unstable and the growth rates
are proportional to the wave vectors, is preserved.

\begin{figure}[h]
\begin{center}
\epsfclipon\includegraphics[angle=0,width=0.9\columnwidth]{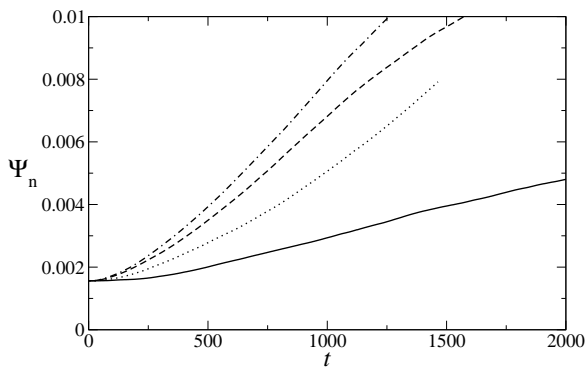}
\caption{Evolution of the Fourier amplitudes $\Psi_n$ computed with the
same wavenumber $n$ as the perturbation made on the surface. $n=1$
continuous line, $n=2$ dotted line, $n=3$ dashed line, and $n=4$
dash-dotted lined¡.}
\label{Fig.parOrden}
\end{center}
\end{figure}

We have studied numerically the growth rate for different wavevectors in
the
jet. For each wavevector we have performed new simulations. with an initial
condition similar to the one described previously for the jet except that
the $x$ coordinates are modulated in such a way to create two waves of
vectors $k_y=2\pi n_y/L_y$ in the surface positions. In practice we map the
initial $x$ coordinates of the original rectangular jet as
$[0,L_x] \to [-A \cos(k_y y),L_x+A
\cos(k_y y)]$. In each simulation we compute the evolution of the Fourier
amplitudes $\Psi_n = N^{-1} |\sum_i (x_i-L_x/2)^2 e^{2\pi i n y_i/L_y} | = 
N^{-1} | \int \rho(x,y) (x-L_x/2)^2 e^{2\pi i n y/L_y} |$, with the same
wavenumber as the perturbation. These Fourier amplitudes measure, at
linear order, the amplitude of $\xi(k_y)$. Fig.
\ref{Fig.parOrden} shows the results of the Fourier
amplitudes for the simulations done with $n_y=1,2,3,4$.
For $n_y=1$ a linear increase of $\Psi$ is observed while for $n_y\geq 2$ 
exponential increase of the $\Psi$'s are obtained, until nonlinear
interaction between modes appear. The growth rates are
$\lambda(n_y=2)=1.21\times10^{-3}$, $\lambda(n_y=3)=1.82\times10^{-3}$, and
$\lambda(n_y=4)=2.19\times10^{-3}$, that
are roughly in the ratio $2:3:4$, confirming the linear proportionality
with $k$. However, no direct comparison with the prediction can be made
because only $k_y$ was fixed but not $k_x$, therefore the angle $\phi$
takes all possible values. Larger values of $n_y$ give very noisy results
to be confident.

\section{Conclusions}

In conclusion, we investigated the dynamical evolution, in the Stokes
regime, of a jet of falling particles confined in an Hele-Shaw cell. The
free surfaces of the jet is shown to become unstable and a coarsening
process takes place from the smallest sizes up to the largest structures.

 A continuous Euler-like hydrodynamic model for the suspension is presented
which describes the main features of the jet flow. In particular, the model
predicts a geometry dependent falling velocity for
the jet (thin jets fall slower than thick jets).

The theoretical analysis shows that the free surface is unstable to any
gravitational force and any wave vector. The growth rates are proportional
to the wave vectors, result that is consistent with the coarsening process
observed in the simulations. Finally, the model predicts that the structures
move with a velocity slightly different than the jet's velocity, as observed
in the simulations but with a different sign prediction. In this problem we
find that the interface dynamics is quite different than the usual
immiscible fluid case where the surface tension play a leading role such as
to select and stabilize structures. For the present jet,the particle
hydrodynamic interactions create a surface instability essentially due to
long range effects caused by flow recirculation around a particle. A crucial
point is that this effect is scale free. We observe at the end of the
simulations a partial mixing due to flow structuration as well as
hydrodynamic dispersion causing a diffuse interface.

Finally, we open a question on the relation between this model and the
dynamics of actual non-Brownian suspensions in a confined cell. Though we
have here a situation a little bit artificial as the problem is solved in
the case of particle sizes (diameter) larger than the cell gap, we believe
that all the features dealing with scales larger than the gap ( essentially
the long range interactions) will survive at the qualitative level.

\begin{acknowledgments}

The authors thank P. Cordero and M. Malek Mansour for their comments.
A.Alvarez and R.Soto. are grateful with the hospitality of CECAM, ENS-Lyon
where part of this work was done. The simulations were done in the CIMAT's
parallel cluster. This work has been partly financed by the {\em Fondecyt}
research grant 1030993, the ECOS-Conicyt research grant C03E05 and the {\em Fondap}
grant 11980002. A.A. acknowledges the financial support of a {\em Mecesup}
grant.

\end{acknowledgments}

\appendix
\section{Kernel integrals}
Some integrals of the kernel $\varmathbb{K}(\vec{R})$ (\ref{kernel})
used in the manuscript are:

\noindent{\bf (a)} Homogenenous integral for the jet
\begin{align}
I_1&=\int_{-L_x}^{L_x} dx' \int_{-L_y}^{L_y} dy'\,
\varmathbb{K}(\vec{R}-\vec{R'}) \nonumber \\ 
&= 2\sigma^2 \left[
\arctan\left(\frac{L_y}{L_x-x}\right)-\arctan\left(\frac{L_x+x}{L_y}
\right)\right ]
\left(\begin{array}{lr} 1 & 0\\ 0 & -1 \end{array} \right)
\end{align}
where $\vec{R}=x\hat{x}+y\hat{y}$ is any
vector in the interior of the integration domain and we have used periodic
boundary conditions in $y$. When $L_y\gg L_y$ the integral depends slightly
on $x$ and its horizontal average is in that limit
\begin{align}
I_1&\approx 2\sigma^2 \left[
\arctan(L_y/L_x)-\arctan(L_x/L_y)\right]
\left(\begin{array}{lr} 1 & 0\\ 0 & -1 \end{array}
\right)\label{IntKFinito}
\end{align}

\noindent{\bf (b)} Homogenenous integral for the semiinfinite system
\begin{align}
I_2&=\int_{-\infty}^{0} dx' \int_{-\infty}^{\infty} dy'\,
\varmathbb{K}(\vec{R}-\vec{R}') =-\frac{\pi\sigma^2}{2} 
\left(\begin{array}{lr} 1 & 0\\ 0 & -1 \end{array} \right)
\label{IntKSemiInf}
\end{align}
with $\vec{R}=x\hat{x}+y\hat{y}$ any vector with $x<0$ and we have
used periodic boundary conditions in $y$.
	
\noindent{\bf (c)} Fourier transform for the semiinfinite
system
\begin{align}
I_3&=\int_{-\infty}^{0} dx'\int_{-\infty}^{\infty} dy'\,
\varmathbb{K}(\vec{R}-\vec{R}') e^{i \vec{k}\cdot\vec{R}'} \nonumber \\
&= \pi\sigma^2 \left[ 
\left(\begin{array}{lr} 1 & 0\\ 0 & -1 \end{array} \right)
-\frac{|k_y|}{ik_x+|k_y|} 
\left(\begin{array}{lr} 1 & -i\\ -i & -1 \end{array} \right)
\right]
\end{align}
with $\vec{R}=x\hat{x}+y\hat{y}$ any vector with $x<0$ and we have
used periodic boundary conditions in $y$.

\noindent{\bf (d)} Fourier transform for the interface line
\begin{align}
I_4&=\int_{-\infty}^{\infty} dy\,
\varmathbb{K}(x\hat{x}+y\hat{y}) e^{i k y} \nonumber \\
&= 
\pi \sigma^2\left\{\delta(x)
 \left(\begin{array}{lr} 1 & 0\\ 0 & -1 \end{array} \right)
-|k| e^{-|kx|} 
\left(\begin{array}{lr} 1 & -i\\ -i & -1 \end{array} \right)
\right\} \label{FourierKy}
\end{align}
The Dirac delta appears when $\varmathbb{K}$ is evaluated up to the
origin, that physically cannot happen because the interaction is replaced
by the lubrications forces. Therefore, this term will be eliminated in the
evaluation of the kernel integrals.


\begin{thebibliography}{100}

\bibitem{Happel65} J. Happel and H. Brenner, ``Low Reynolds Number Hydrodynamics, with
Special Applications to Particulate Media'' Prentice-Hall, London, 1965.

\bibitem{refe_SD} J.F. Brady and G. Bossis, 
``Stokesian dynamics'', Annu. Rev. Fluid. Mech. \textbf{20}, 111 (1988).

\bibitem{Ladd02}A. J. C. Ladd, ``Effect of container walls on the velocity fluctuations of
sedimenting spheres'' Phys. Rev. Lett. \textbf{88}, 048301 (2002).

\bibitem{lubrica} S. Kim and S. J. Karrila, {\em Microhydrodynamics:
Principles and Selected Applications} Butterworth-Heinemann Series in
Chemical Engineering, Stoneham 1991.

\bibitem{Stone04} H.A.Stone, A.D.Stroock and A.Ajdari, 
``Engineering flows in small devices : Microfluidics towards a lab-on-chip'', Annu. Rev. Fluid. Mech. \textbf{20}, 111 (2004).

\bibitem{Weigl99} B.H.Weigl and P. Yager, 
``Microfluidic diffusion-based separation and detection'', Science \textbf{283}, 346 (1999).


\bibitem{Rouyer1999} F. Rouyer, J. Martin, and D. Salin, ``Structure,
density, and velocity fluctuations in quasi-twodimensional non-Brownian
suspensions of spheres'', Phys. Rev. Lett.
\textbf{83}, 1058 (1999).

\bibitem{Santana01} J. Santana-Solano and J. Arauz-Lara, ``Hydrodynamic interactions in
quasi-two-dimensional colloidal suspensions'', Phys. Rev. Lett.
\textbf{87}, 038302 (2001).

\bibitem{cui2004} B. Cui et al., ``Anomalous hydrodynamic interaction in a quasi-twodimensional
suspension'' Phys. Rev. Lett. \textbf{92}, 258301 (2004).

\bibitem{petitjean99}P. Petitjeans, C.Y. Chen, E. Meiburg, T. Maxworthy
``Miscible quarter five-spot displacements in a Hele-Shaw cell and the role
of flow-induced dispersion'' Phys. Fluids, \textbf{11}1705 (1999).

\bibitem{refe8} J. M. Nitsche and G. K. Batchelor, ``Break-up of a falling drop containing
dispersed particles'', J. Fluid Mech.,
\textbf{340}, 161 (1997).

\bibitem{refe11} G. Machu, W. Meile, L. C. Nitsche and U. Schaflinger,
``Coalescence, torus formation and breakup of sedimenting drops: experiments and computer
simulations'' , J. Fluid Mech., \textbf{447}, 299 (2001).

\bibitem{refe9} Maxime Nicolas,  Phys. Fluids {\bf 14}, ``Experimental study
of a gravity driven dens suspension jet'' Phys. of Fluids {\bf 14}, 3570
(2002).
 
\bibitem{hele_suspe} A. Alvarez and R. Soto, Phys. of Fluids ``Dynamics of a
suspension confined in a thin cell'' Phys. of Fluids {\bf 17}, 093103
(2005).

\bibitem{hele_suspe_jet} A. Alvarez R. Soto and E.Clement, ``Free surface
instability in a confined suspension jet'', Physica A {\bf 356}, 196
(2005).

\bibitem{Allen}
M.~P. Allen y D.~J. Tildesley, {\em Computer Simulation of Liquids\/} (Oxford
  Science Publications, New York, 1990).
  
\bibitem{chandra} S. Chandrasekhar, {\em Hydrodynamic and Hydromagnetic
Stability}. International Series of Monographs on Physics,  Dover, Oxford 
1981.


\bibitem{KT} S. Chapman and T. G. Cowling, {\em The Mathematical
Theory of Non-Uniform Gases}, third edition. Cambridge University Press,
New York, 1970.

\bibitem{coarsening} P. Politi and C. Misbah,``When Does Coarsening Occur in
the Dynamics of One-Dimensional Fronts?''
 Phys. Rev. Lett. \textbf{92}, 090601 (2004)

 
\end{thebibliography}
\end{document}